\documentclass[conference,letterpaper,twoside,10pt]{IEEEtran}
 
\usepackage[pdftex]{graphicx}
\usepackage[cmex10]{amsmath}
\usepackage[]{algorithm2e}
\usepackage{url}
\usepackage{color}
\usepackage{amssymb}
\usepackage{cite}
\usepackage{bm}
\usepackage{subcaption}
\usepackage{lipsum}
\usepackage{epstopdf}
\usepackage{physics}
\usepackage{fancyhdr}
\usepackage{comment}

\usepackage[cmex10]{amsmath,empheq}

\makeatother

\begin{document}
\title{Digital Cancellation of Passive Intermodulation in FDD Transceivers}

\author{\IEEEauthorblockN{Muhammad Zeeshan Waheed, Pablo Pascual Campo, Dani Korpi, Adnan Kiayani, Lauri Anttila, Mikko Valkama}
\\[-5pt]
\IEEEauthorblockA{Laboratory of Electronics and Communications Engineering, Tampere University of Technology, Finland\\e-mail: muhammad.waheed@tut.fi}
}

\maketitle

\begin{abstract}
Modern radio systems and transceivers utilize carrier aggregation (CA) to meet the demands for higher and higher data rates. However, the adoption of CA in the existing Long Term Evolution (LTE)-Advanced and emerging 5G New Radio (NR) mobile networks, in case of frequency division duplexing (FDD), may incur self-interference challenges with certain band combinations. More specifically, the nonlinear distortion products of the transmit signals or component carriers (CCs), stemming from the passive radio frequency (RF) front-end components of the transceiver, can appear in one or more of the configured receiver bands, potentially leading to the receiver desensitization. In this paper, we present advanced baseband equivalent signal models for such passive intermodulation (PIM) distortion viewed from the RX point of view, considering also potential memory effects in the PIM generation. Then, building on these signal models, a digital self-interference cancellation technique operating in the transceiver digital front-end is presented. The performance of the proposed solution is evaluated with real-life RF measurements for LTE-Advanced type user equipment (UE) with dual CC inter-band CA, demonstrating excellent suppression properties. The findings in this work indicate that digital cancellation is a feasible approach for improving the receiver sensitivity of mobile devices that may be prone to RF front-end induced PIM challenges.
\end{abstract}

\vspace{2mm}
\begin{IEEEkeywords}
4G LTE-Advanced, 5G NR, digital cancellation, frequency division duplexing, nonlinear distortion, passive intermodulation, self-interference.
\end{IEEEkeywords}

\IEEEpeerreviewmaketitle

\section{Introduction}
The increasing amounts of data usage by mobile network subscribers imply the need for higher throughputs and higher network capacities. The existing and emerging mobile communication networks, such as 4G long term evolution (LTE)-Advanced and 5G New Radio (NR), are designed to meet these needs and requirements \cite{DAHLMAN201857}. Carrier aggregation (CA) is one of the key techniques that was introduced in LTE-Advanced to support higher throughput requirements, where multiple component carriers (CCs) at one or multiple LTE bands are aggregated together to form larger transmission bandwidth, and also to facilitate efficient utilization of the available radio spectrum \cite{3GPP_36850, LTE, Iwamura}. In this work, we particularly focus on the case where the aggregated CCs are at different bands, commonly referred to as inter-band CA.

In general, modern radio systems employing wideband multicarrier waveforms are vulnerable to practical analog circuit implementation related challenges and imperfections. One of these challenges is the so called passive intermodulation (PIM) that can severely limit the performance of frequency division duplexing (FDD) based systems. Such PIM is typically generated in the passive components of the radio frequency (RF) transceiver front-end, such as duplexer, diplexer, multiplexer or antenna selection switches. Moreover, the nonlinear junctions typically caused by poor RF connection or the presence of dirt over the metal surfaces in the radio components can also generate PIM. As a consequence, unwanted nonlinear distortion products are created due to the intermodulation of the transmit CCs that generally appear at the specific intermodulation (IM) sub-bands. Depending on the used bands and adopted CC center-frequencies, some of these nonlinear PIM products may appear in one or more of the receiver operating bands as illustrated in Fig. 1. Furthermore, since the PIM is generated in the radio transceiver front-end at or after the duplexer filter, it leaks directly into the receiver and may lead to receiver desensitization. 

\begin{figure}[t]
\centering
\includegraphics[scale=0.4]{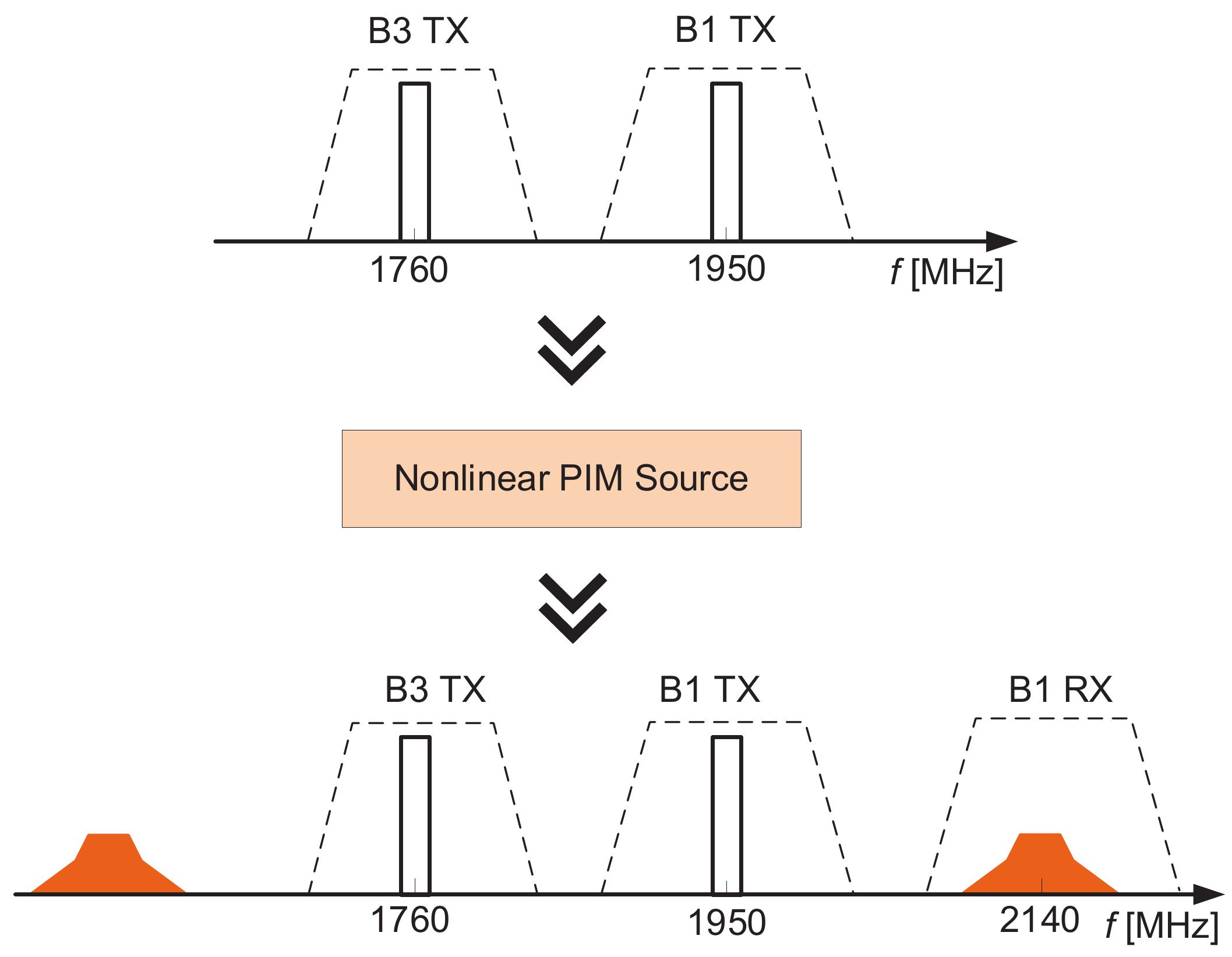}
\caption{{Spectral illustration of the unwanted PIM products with interband CA of Band 1 and Band 3 at mobile device side. In this example, some of the PIM products hit the Band 1 receiver.}}\label{spectra}
\end{figure}

A concrete example case, in terms of exact LTE bands and frequencies, is given in Fig.~\ref{spectra} illustrating uplink inter-band CA transmission at Band 1 (1920-1980 MHz) and Band 3 (1710-1780 MHz). As shown in the figure, the upper third-order IM sub-band (IM3) falls within the Band 1 downlink, reflecting thus the self-interference problem due to PIM. Other LTE bands that can experience similar problems are, e.g., B3+B8, B2+B4, B5+B7, as discussed and acknowledged also in many inter-band CA related 3GPP technical documents, such as \cite{3GPP_36101}, \cite{121131}. In general, the problem of PIM-induced self-interference is not only limited to UE devices but can actually be even more pronounced in the base station (BS) transceiver systems \cite{1418994}, \cite{8077999} where, in addition to internal PIM sources, external sources such as metal objects in the antenna near field and reflections from nearby buildings can cause self-interference. Therefore, PIM can be a big concern also for network vendors and operators. 

Obvious solutions to avoid or reduce the PIM-induced self-interference are to either reduce the transmit signal power or to allow for a degradation in the receiver reference sensitivity level. At the UE side, these approaches are referred to as the maximum power reduction (MPR) and maximum sensitivity degradation (MSD), respectively. However, these approaches impact negatively the UL link budget and throughputs \cite{Adnan_MTT} and are thus not the most appealing solutions. Alternatively, one could argue to utilize higher quality RF components with good linearity characteristics, however, this may considerably raise the overall radio implementation costs and size.

Some recent works have addressed digital cancellation of PIM. Specifically, Dabag et al. \cite{Dabag_2} considered third-order PIM cancellation caused by the antenna switch by devising a multiple input single output (MISO) canceller to suppress the frequency-selective PIM with time delay differences between different transmit signals. While showing promising results, the associated parameter estimation complexity is very high. Then, in \cite{Dabag_1}, digital cancellation of second-order PIM due to a diplexer is pursued. In general, these reference techniques do not take into account the nonlinear behavior of the individual PAs in the transmitter chains and the memory effects of the PAs. Since the transmit CCs are distorted in a nonlinear fashion by each of the individual PAs before entering a PIM nonlinearity, this implies that the self-interference is in fact a combination of two nonlinearities. This has been identified recently in \cite{zeeshan_SIPS} and \cite{Zeeshan_MTT_submit}, where different digital cancellers for joint mitigation of PA and PIM nonlinerities are proposed and experimented. 

In this paper, we develop advanced digital cancellation solutions for suppressing the PIM with memory effects while exclude the PA nonlinearities for simplicity. In addition to PA memory, the proposed method can also account for different mutual time delays of the transmit signals before entering the PIM source, while can also account for memory along and after the PIM generation stage. For presentation simplicity, we focus primarily on modeling and digital cancellation of third-order PIM. 
The performance of the developed method is evaluated through practical RF measurements, adopting commercial LTE/LTE-Advanced UE transceiver modules and RF components. Moreover, we also address in the performance evaluations and RF measurements an additional practical case 
where the UE is equipped with a diversity RX chain. Specifically, we show that \emph{PIM coupled over-the-air} from main transceiver to the diversity RX can also be a real problem. Although such PIM coupling over the air can basically be avoided by improving the isolation between the antennas, the isolation is in practice limited by the compact size of the mobile devices. The proposed digital cancellation solution is shown to be able to efficiently suppress the PIM appearing in the main RX branch as well as in the diversity RX branch. Thus, in general, the proposed solution can relax the RF components' linearity requirements and improve the receiver sensitivity by effectively suppressing the PIM, and is applicable in both main RX and diversity RX branches. 

The rest of the paper is organized as follows. In Section II, we address baseband equivalent modeling of third-order PIM at RX band under various sources of memory. The corresponding digital cancellation solution and the necessary parameter estimation procedures are presented in Section III. Then, the RF measurement results are reported and analyzed in Section IV. Finally, Section V concludes the paper.

\begin{figure}[t]
\centering
\includegraphics[scale=0.42]{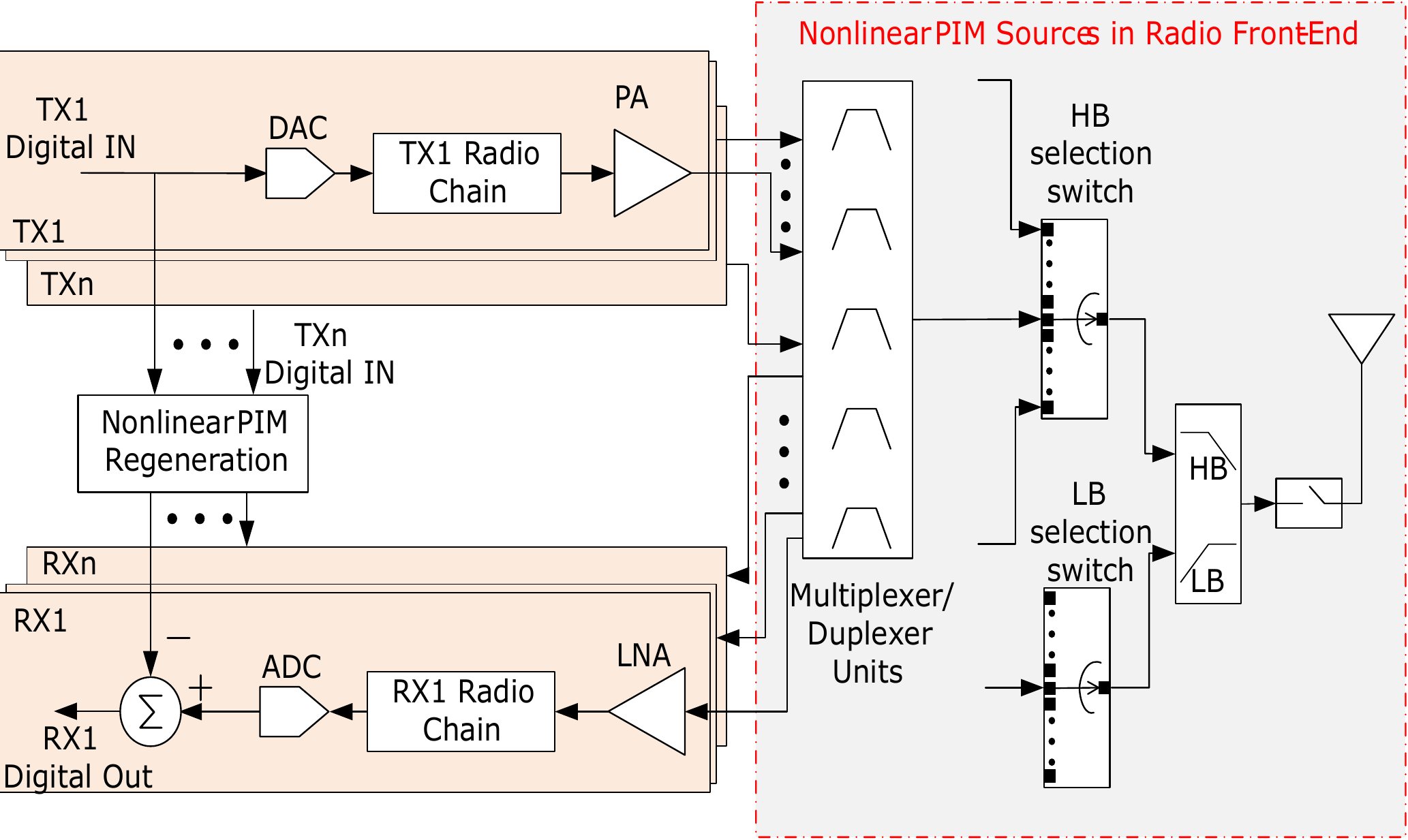}
\caption{{The considered inter-band CA FDD transceiver architecture at UE side.}}\label{trans_architecture}
\end{figure}

\section{Baseband Equivalent Models for Passive Intermodulation at RX Band}
In this section, we present the relevant signal models for describing the PIM observed in the RX chain, particularly at digital baseband. A block diagram describing the considered FDD radio transceiver system supporting inter-band CA is shown in Fig.~\ref{trans_architecture}. It is assumed that the adopted bands have dedicated TX-RX chains, each, while on the transmit direction the CCs are combined together in a duplexer or a multiplexer \cite{121984}. The PIM then occurs in the passive RF components due to cross-modulation between the aggregated transmit signals, and may appear in one or more of the configured receiver bands causing nonlinear self-interference. 

Generally speaking, as shown in \cite{zeeshan_SIPS} and \cite{Zeeshan_MTT_submit}, the TX PAs can also cause nonlinear distortion to the individual transmit carriers, leading to spectral regrowth around the main transmit carriers. Such nonlinear distortion in the transmit CCs can then affect the overall characteristics of the PIM-induced self-interference and its cancellation. However, in this paper, we restrict our attention to linear transmit chains and PAs, thus basically assuming that the PAs are properly linearized, e.g., through digital pre-distortion (DPD). It can, however, be argued that the transmit carriers experience some filtering or linear distortion effects before they are combined and thus experience the PIM nonlinearity. Additionally, there can also potentially be multiple PIM sources in the transceiver. As a result, the overall TX-RX PIM coupling path can be viewed as a system of nonlinearities with memory effects. Keeping this in view, we consider two cases for PIM modeling. In the first case, we consider linear and memoryless PAs and pursue frequency-selective behavioral modeling of the nonlinear passive components. In the second case, we also adopt the memory or linear distortion modeling of the individual transmit CCs,  prior to the actual PIM stage. As we later demonstrate with practical RF measurements, the latter model can facilitate more accurate PIM modeling and enhanced cancellation.

For presentation purposes and notational simplicity, we focus on third-order PIM effects, while more elaborate higher-order cases as well as coexisting cascaded nonlinearities are addressed in \cite{Zeeshan_MTT_submit}.

\subsection{Baseband Equivalent PIM Model: Memoryless TX Chains}\label{sec:modelA}
Let us denote the complex baseband waveforms of the two transmit CCs 
by $s_1[n]$ and $s_2[n]$, respectively. The signals are up-converted to their respective RF frequencies and amplified by the PAs. The corresponding aggregated RF signal at the multiplexer/duplexer output, after combining the carriers, can then be expressed as
\begin{equation}\label{TXsignal}
	s_{\mathrm{RF}}[n] = \Re{\alpha_1s_1[n] e^{j \omega_1 n}} + \Re{\alpha_2s_2[n] e^{j\omega_2 n}},
\end{equation}
where $\alpha_1$ and $\alpha_2$ denote the complex voltage gains of the two PAs, and $\omega_1$ and $\omega_2 < \omega_1$ are the angular center frequencies of the individual CCs after up-conversion. This signal then travels towards the antenna, however, due to nonlinear passive components, unwanted PIM products of the transmit signal are created. Assuming that the upper third-order IM sub-band, i.e. $2\omega_1-\omega_2$, lies in the downlink frequency band, similar to Fig. 1, the baseband equivalent complex PIM waveform appearing in the RX band reads then 
\begin{equation}\label{PIM}
	s_{\mathrm{PIM}}[n] = \sum_{l=-L_1}^{L_2} \gamma_l s_1[n-l]^2s_2^*[n-l] 
\end{equation}
where $\gamma_l$ denote the impulse response coefficients of the third-order nonlinear term modeling the memory of the PIM generation mechanism, while $L_1$ and $L_2$ are the numbers of pre-cursor and post-cursor memory taps in the PIM model, respectively. The total memory length of the PIM generation stage is then $L_1 + L_2 + 1$.

\subsection{Baseband Equivalent PIM Model: TX Chains with Memory}\label{sec:modelB}
We next generalize the PIM modeling to the case where the individual TX component carrier signals are subject to memory or linear distortion prior to the PIM stage. 
When complemented with a memory polynomial model for the actual PIM generation stage, this allows for versatile memory modeling, for example in cases where there are mutually different delays along the TX paths, or more generally the frequency responses of the two TX chain are different and both contain memory. 

To this end, the two transmit carriers travel through their independent TX chains before arriving at the PIM source, and are subject to linear filtering effects described by the impulse responses $\alpha_{1,m}$ and $\alpha_{2,m}$, respectively. 
%
%
The corresponding RF signal model for the combined signal, prior to the PIM stage, then reads
\begin{equation}\label{TXsignal2}
\begin{split}
	s_{\mathrm{RF}}[n] = \Re{e^{j \omega_1 n}\sum_{m=-M_1}^{M_2} \alpha_{1,m}s_1[n-m]} \\
	+ \Re{e^{j\omega_2 n}\sum_{m=-M_1}^{M_2} \alpha_{2,m}s_1[n-m]},
\end{split}
\end{equation}
where $M_1, M_2$ are the numbers of the input pre- and post-cursor memory taps for the TX carriers, with the total input memory length being $M_1+M_2+1$. 
When the above combined signal with memory is subject to a third-order PIM nonlinearity with additional memory, the baseband equivalent PIM waveform at own RX band can be written as shown in \eqref{equation_mem_PIM}, next page, where $\gamma_{k_{11}, \cdots, k_{1M}, k_{21}, \cdots, k_{2M}}$ are the effective total memory coefficients for the term defined by the parameters $k_{1M}, k_{21}, \cdots, k_{2M}$, and $M = M_1+M_2$. Note that in this model, the memory lengths of the transmit CCs and the PIM nonlinearity can be independently chosen, resulting in an overall flexible model from the modeling and cancellation perspective.

To give a concrete example, Table \ref{table:Basis} shows the basis function samples stemming from the two signal models presented in this section, with short memory orders of $L_1=L_2=1$ and $M_1=M_2=1$ for presentation simplicity. As can be observed, the signal model under TX chains with memory has altogether $42$ basis functions, opposed to $3$ basis functions obtained in the memoryless TX chain case. As a consequence of the different sample delays between the CCs in the basis functions, the more complicated signal model has the potential of better estimating and cancelling also the impacts of any potential timing mismatch errors as well as overall frequency responses in the TX chains. This is achieved, however, at the cost of an increased computational complexity. 
\begin{figure*}[ht]
\begin{equation}\begin{split}\label{equation_mem_PIM}
    s_{\mathrm{PIM}}[n] = &\sum_{k_{11}=0}^{2}\sum_{k_{12}=0}^{2-k_{11}}\cdots\sum_{k_{1M}=0}^{2-\sum_{i=1}^{M-1}k_{1i}}
\sum_{k_{21}=0}^{1}\sum_{k_{22}=0}^{1-k_{21}}\cdots\sum_{k_{2M}=0}^{1-\sum_{i=1}^{M-1}k_{2i}} \sum_{l=-L_1}^{L_2}\gamma_{l,k_{11}, \cdots, k_{1M}, k_{21}, \cdots, k_{2M}} \times \\
&s_1[n-l+M_1]^{k_{11}}s_1[n-l+M_1-1]^{k_{12}}\cdots s_1[n-l-M_2]^{2-\sum_{i=1}^{M}k_{1i}} \times \\ 
&s_2^*[n-l+M_1]^{k_{21}}s_2^*[n-l+M_1-1]^{k_{22}}\cdots s_2^*[n-l-M_2]^{1-\sum_{i=1}^{M}k_{2i}}.
\end{split}
\end{equation}
\vspace{2mm}
\hrulefill
\end{figure*}
\setlength{\arrayrulewidth}{0.2mm}
\setlength{\tabcolsep}{3pt}
\renewcommand{\arraystretch}{1.5}
%
\begin{table*}[h]
\caption{Example basis functions of the two considered PIM models}
\label{table:Basis}
\centering
\begin{tabular}{ | p{3cm} | p{14cm} | }
\hline
 \textbf{Signal model}  & \textbf{Basis functions when $L_1=L_2=1$} \\
 \hline
 Memoryless TX Chains & {\centering{}$s_1[n+1]^2 s_2^*[n+1]$, $s_1[n]^2 s_2^*[n]$, $s_1[n-1]^2 s_2^*[n-1]$\\ \vspace{-10pt}}\\
 \hline
 TX Chains with Memory, $M_1=M_2=1$ & {\centering{} $s_1[n-2]^2 s_2^*[n-2]$, $s_1[n-1]^2 s_2^*[n-1]$, $s_1[n]^2 s_2^*[n]$, $s_1[n-2]^2 s_2^*[n-1]$, $s_1[n-1]^2 s_2^*[n]$, $s_1[n]^2 s_2^*[n+1]$,\\$s_1[n-2]^2 s_2^*[n]$, $s_1[n-1]^2 s_2^*[n+1]$,$s_1[n]^2 s_2^*[n+2]$, $s_1[n-1] s_1[n-2] s_2^*[n-2]$, $s_1[n] s_1[n-1] s_2^*[n-1]$,\\$s_1[n+1] s_1[n] s_2^*[n]$, $s_1[n-1] s_1[n-2] s_2^*[n-1]$, $s_1[n] s_1[n-1] s_2^*[n]$, $s_1[n+1] s_1[n] s_2^*[n+1]$,\\$s_1[n-1] s_1[n-2] s_2^*[n]$, $s_1[n] s_1[n-1] s_2^*[n+1]$, $s_1[n+1] s_1[n] s_2^*[n+2]$, $s_1[n] s_1[n-2] s_2^*[n-2]$,\\$s_1[n+1] s_1[n-1] s_2^*[n-1]$, $s_1[n+2] s_1[n] s_2^*[n]$, $s_1[n] s_1[n-2] s_2^*[n-1]$, $s_1[n+1] s_1[n-1] s_2^*[n]$,\\$s_1[n+2] s_1[n] s_2^*[n+1]$, $s_1[n] s_1[n-2] s_2^*[n]$, $s_1[n+1] s_1[n-1] s_2^*[n+1]$, $s_1[n+2] s_1[n] s_2^*[n+2]$,\\$s_1[n] s_1[n-1] s_2^*[n-2]$, $s_1[n+1] s_1[n] s_2^*[n-1]$, $s_1[n+2] s_1[n+1] s_2^*[n]$, $s_1[n+2] s_1[n+1] s_2^*[n+2]$,\\$s_1[n+2]^2 s_2^*[n+1]$, $s_1[n+2]^2 s_2^*[n+2]$, $s_1[n]^2 s_2^*[n-2]$, $s_1[n+1]^2 s_2^*[n-1]$, $s_1[n+2]^2 s_2^*[n]$,\\$s_1[n-1]^2 s_2^*[n-2]$, $s_1[n]^2 s_2^*[n-1]$, $s_1[n+1]^2 s_2^*[n]$, $s_1[n+1]^2 s_2^*[n+1]$, $s_1[n+1]^2 s_2^*[n+2]$,\\$s_1[n+2] s_1[n+1] s_2^*[n+1]$\\ \vspace{-10pt}}\\
 \hline
 \end{tabular}
 \end{table*}
 

\section{Proposed Digital PIM Canceller and Parameter Estimation}
The proposed digital PIM canceller builds directly on the derived baseband signal models described in the previous section. Specifically, the canceller re-generates the PIM samples, using either (\ref{PIM}) or (\ref{equation_mem_PIM}), and then subtracts the estimated PIM samples from the actual received baseband signal. In general, the baseband complex samples of the component carriers, $s_1[n]$ and $s_2[n]$, are known in the transceiver. However, the equivalent model parameters, i.e., the $\gamma$ variables 
which act as the complex weights of the basis function samples are unknown and thus must be estimated.
%

Regarding the parameter estimation, the baseband PIM signal models in (\ref{PIM}) and (\ref{equation_mem_PIM}), and thus the canceller processing structures, are in fact linear in the parameters, i.e., the $\gamma$ variables. 
Thus, the parameters can be straight-forwardly estimated with any standard estimator for linear signal models, such as linear least-squares (LS), recursive least squares (RLS), or least mean squares (LMS) \cite{Haykin}.

For presentation purposes, we next switch to vector-matrix notations and consider a block of $N$ samples of the received signal. We express the samples of the PIM for the corresponding time duration, stacked into a vector, as
\begin{align}
	\mathbf{s}_{\mathrm{PIM}} = \bm{A} \bm{\theta},
\end{align}
where $\bm{A}$ denotes the data matrix that collects the samples of the different nonlinear basis functions while $\bm{\theta}$ is a vector containing the corresponding unknown coefficients. The structure of the data matrix $\bm{A}$ depends obviously on which of the two canceller structures is deployed. Using these notations, the LS parameter estimator reads 
\begin{align}
\widehat{\bm{\theta}} = \left(\bm{A}^H \bm{A}\right)^{-1} \bm{A}^H\mathbf{y}_{\mathrm{RX}},
\label{eq:alpha_estimates}
\end{align}
where $(.)^H$ denotes the Hermitian transpose, and $\mathbf{y}_{\mathrm{RX}}$ is the vector of the received complex baseband samples. 

After obtaining the estimates of the coefficients, they are used to create a baseband replica of the PIM-induced self-interference in the digital front-end of the radio, during the actual online operation, which is subtracted from the received signal sample by sample. 
The cancelled signal during the transceiver online operation can thus formally be expressed as
\begin{align}
    y_\mathrm{RX,canc}[n] = y_{\mathrm{RX}}[n] - \bm{a}^T[n] \hat{\bm{\theta}},
\end{align}
where $\bm{a}^T[n]$ refers to the row vector containing the basis function samples corresponding to the particular time instant $n$.

In general, the basic approach as described above is that the parameter estimation is carried out offline. In such operating approach, the parameter estimation must then be repeated periodically as the exact PIM characteristics can  change over time, e.g., due to changes in the temperature or the transmit power. However, since the PIM is primarily caused by radio transceiver internal front-end components, it is likely that the estimation needs to be repeated only fairly seldom. 
Alternatively, recursive least squares type of adaptive parameter estimation and tracking approach can also be deployed meaning that the parameters are continuously adapted, sample by sample, during the normal online operation. In this case, it is to be noted that the actual received signal acts as noise from the parameter estimation point of view. In both cases, periodic or continuous parameter estimation, the actual cancellation is anyway running continuously in real-time, using the online transmit data.
\begin{figure}[t]
\centering
\includegraphics[scale=0.42]{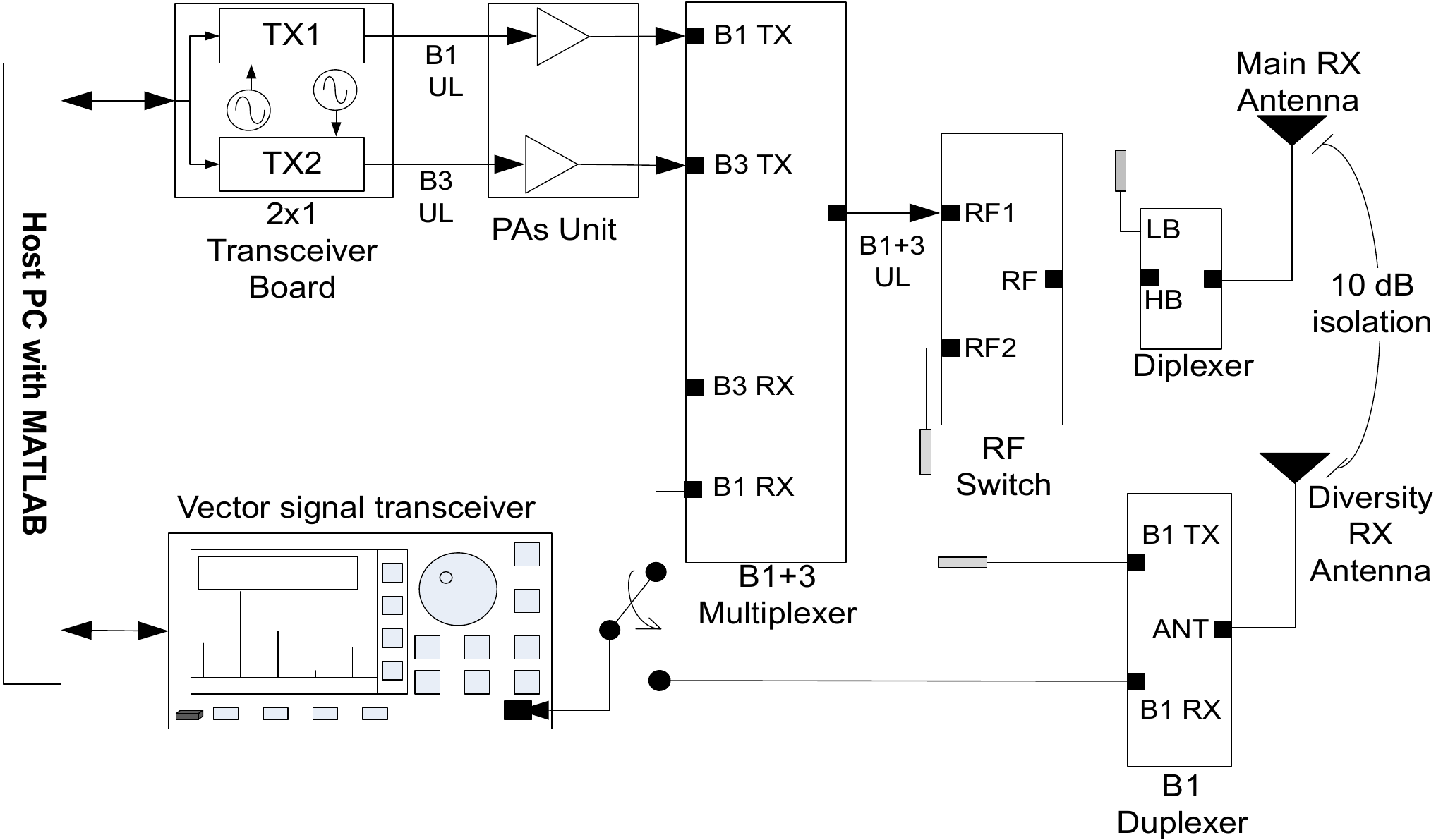}
\caption{{Illustration of the RF measurement setup, showing the relevant transmitter and receiver units as well as the RF front-end components.}}
\label{meas_setup}
\end{figure}

\section{RF Measurement Results}
\subsection{Measurement Setup and Settings}
The performance of the proposed PIM cancellation method is now evaluated through practical RF measurements. A block diagram of the corresponding measurement setup is shown in Fig.~\ref{meas_setup}, while the relevant measurement parameters are listed in Table~\ref{table:simu}.
\begin{table}[t]
\setlength{\tabcolsep}{5pt}
\renewcommand{\arraystretch}{1.3}
\caption{Basic RF measurement parameters}
\label{table:simu}
\centering
\begin{tabular}{|c||c|}
\hline
\textbf{Parameter} & \textbf{Value}\\
\hline
Bandwidths of the TX CCs & 5 MHz\\
\hline
Total transmit power & 24 dBm\\
\hline
Post PA loss & 4 dB\\
\hline
Duplexer insertion loss & 3 dB\\
\hline
Switch insertion loss & 1 dB\\
\hline
RX center frequency & 2140 MHz\\
\hline
LS parameter learning sample size & 90000\\
\hline
Number of PIM pre-cursor taps $(L_1)$ & 3\\
\hline
Number of PIM post-cursor taps $(L_2)$ & 4\\
\hline
Number of TX/PA pre-cursor taps $(M_1)$ & 0 or 1\\
\hline
Number of TX/PA post-cursor taps $(M_2)$ & 0 or 1\\
\hline\end{tabular}
\end{table}
 The measurement setup includes the Analog Devices AD9368 $2\times 1$ transceiver board to generate the Band 1 and Band 3 CC transmit signals. These signals are next amplified using two separate Skyworks SKY77643-21 PAs and combined in a multiplexer TDK B8690, which has separate TX and RX ports and a common antenna port. The aggregated TX signal is then fed to an Infineon BGS12PL6 switch and a TDK DPX162690DT-8022B2 diplexer. The diplexer output port is connected to an antenna. In addition, to implement a diversity RX chain, an additional B1 duplexer connected to an additional antenna is placed in close proximity of the main transceiver. The signals at the B1 RX ports are fed to the RF input of the National Instrument (NI) PXIe-5645R vector signal transceiver (VST), which is used here for down-conversion and digitization of the received signals for further processing. A host processor with MATLAB is used for post-processing the captured data and for the proposed algorithm evaluation. Block least-squares is used for parameter learning, without actual RX signal. Different sets of transmit signal samples are always used, for parameter learning and for evaluating the cancellation performance. All the measurements are performed in an electromagnetic compatibility (EMC) chamber to avoid any external interferences that might affect the results.
\subsection{Measurement Results for Main RX Branch}
\begin{figure}[t]
\centering
\includegraphics[scale=0.5]{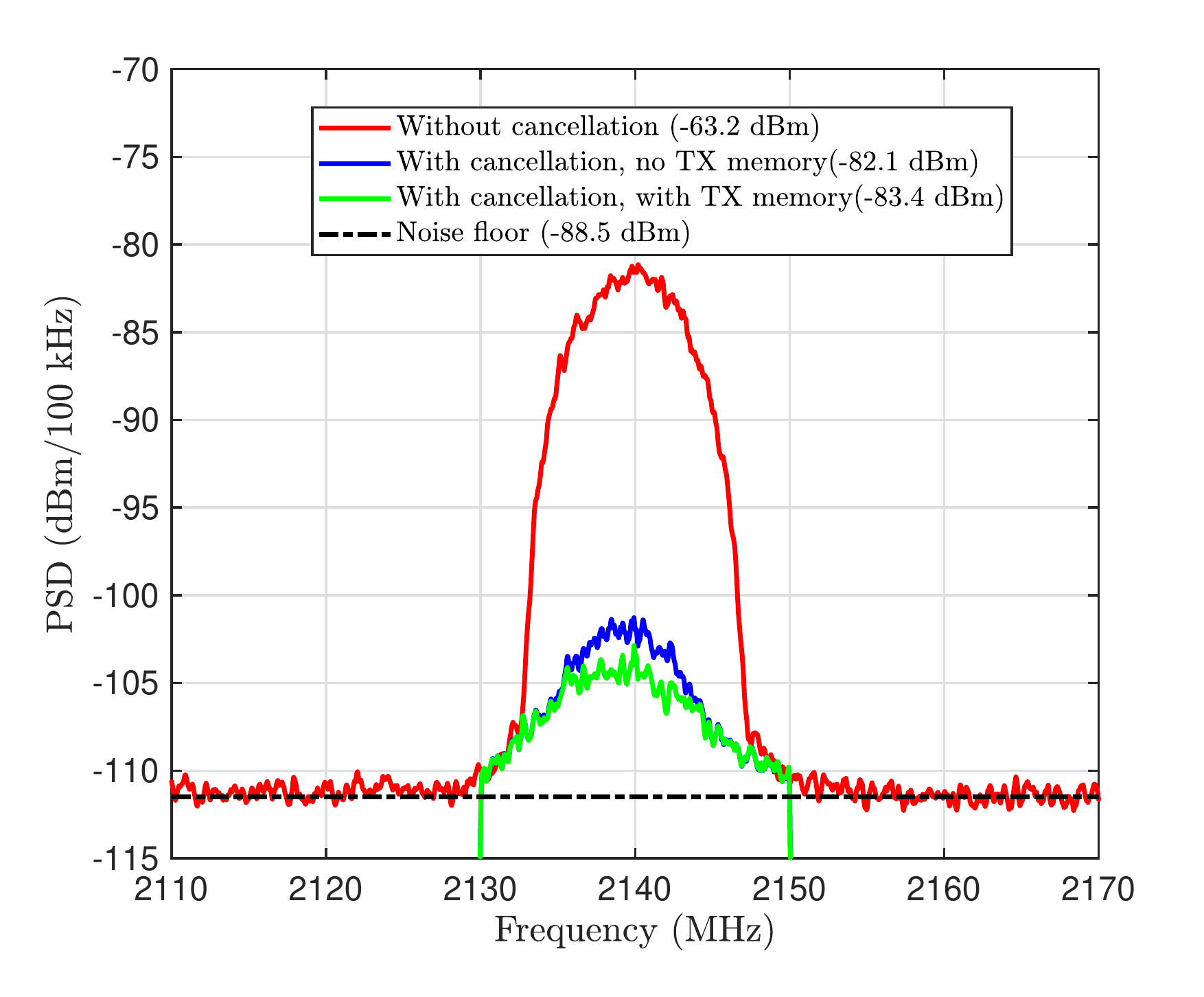}
\caption{{Measured PSD curves of the PIM distortion at the main RX branch of Band 1 without and with digital cancellation. The TX CC bandwidths at the LTE Bands 1 and 3 are $5\;$MHz each, and the aggregated TX power is +$24\;$dBm.}}
\label{Main_RX_PSD}
\end{figure}

In this section, we first present the cancellation results for the main RX branch. The bandwidths of the uplink CCs are assumed to be 5 MHz. When adopting the full transmit power of $+24$ dBm, the essential power spectral density (PSD) curves of the observed PIM before and after the digital cancellation are shown in Fig.~\ref{Main_RX_PSD}, containing both canceller cases of with and without memory in the TX chains. Notice that for illustration purposes, the PSD curves are all referenced to the actual receiver RF frequencies despite the digital canceller operates at baseband. For one, these results clearly indicate that the PIM induced self-interference is significantly above the receiver noise floor when employing state-of-the-art radio components, and can thus cause RX desensitization.
%
\begin{figure}[t]
\centering
\includegraphics[scale=0.5]{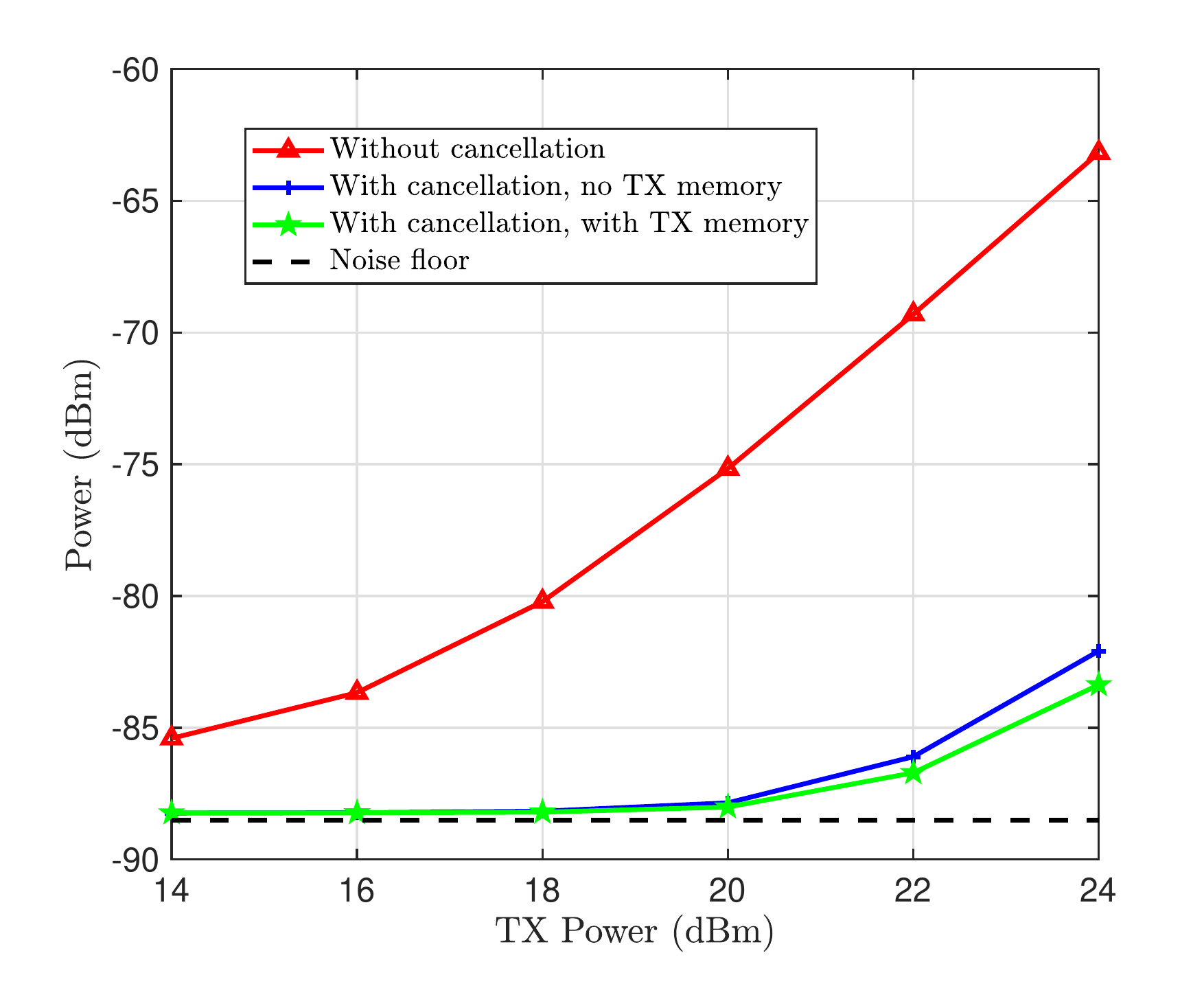}
\caption{{The measured powers of the PIM distortion at the main RX branch of Band 1 without and with digital cancellation as functions of the aggregated transmit power. The TX CC bandwidths at the LTE Bands 1 and 3 are $5\;$MHz each.}}
\label{CancCurves_MainRX}
\end{figure}
However, the proposed digital cancellation solutions are able to suppress the PIM induced self-interference by up to 21 dB or so. Moreover, it can also be noticed that the memory modeling of the PAs, or TX chains overall, prior to the PIM generation stage can further improve the cancellation performance by ca. 1 dB.
 
Next, Fig.~\ref{CancCurves_MainRX} shows the behavior of the PIM-induced self-interference power at Band 1 main RX as a function of the TX power, without and with digital cancellation. The results suggest that the self-interference is already significant even at lower TX powers of some $+14\;$dBm, and can heavily degrade the system performance as the TX power is increased. It can also be observed that the proposed cancellation approach can guarantee interference-free reception for the TX powers up to $+20\;$dBm, thus substantially extending the usable transmit power range without the risk of receiver desensitization.
\subsection{Measurement Results for Diversity RX Branch}
\begin{figure}[t]
\centering
\includegraphics[scale=0.5]{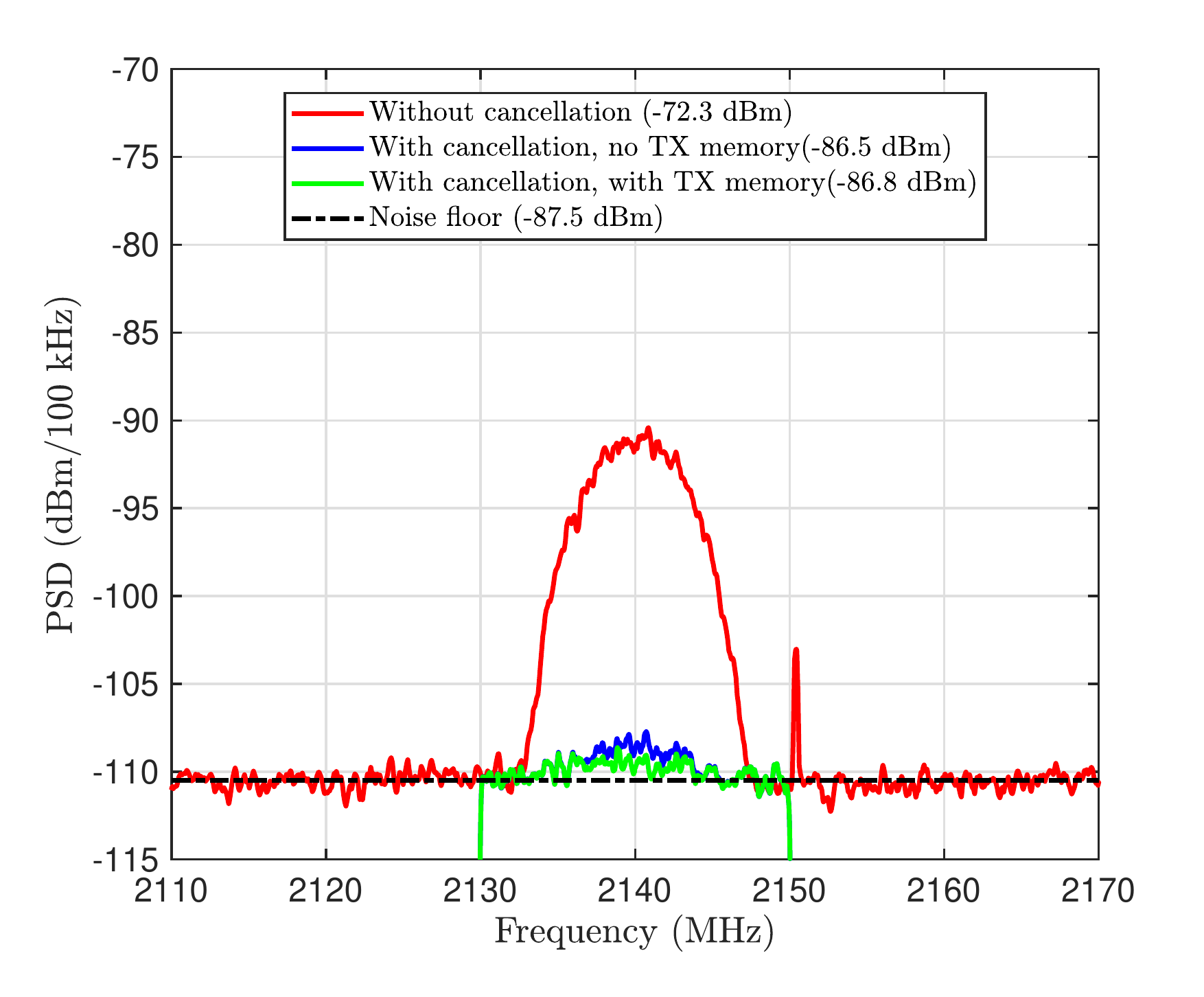}
\caption{{
Measured PSD curves of the PIM distortion at the diversity RX branch of Band 1 without and with digital cancellation. The TX CC bandwidths at the LTE Bands 1 and 3 are $5\;$MHz each, and the aggregated TX power is +$24\;$dBm.
}}
\label{diversity_Rx}
\end{figure}

\begin{figure}[t]
\centering
\includegraphics[scale=0.5]{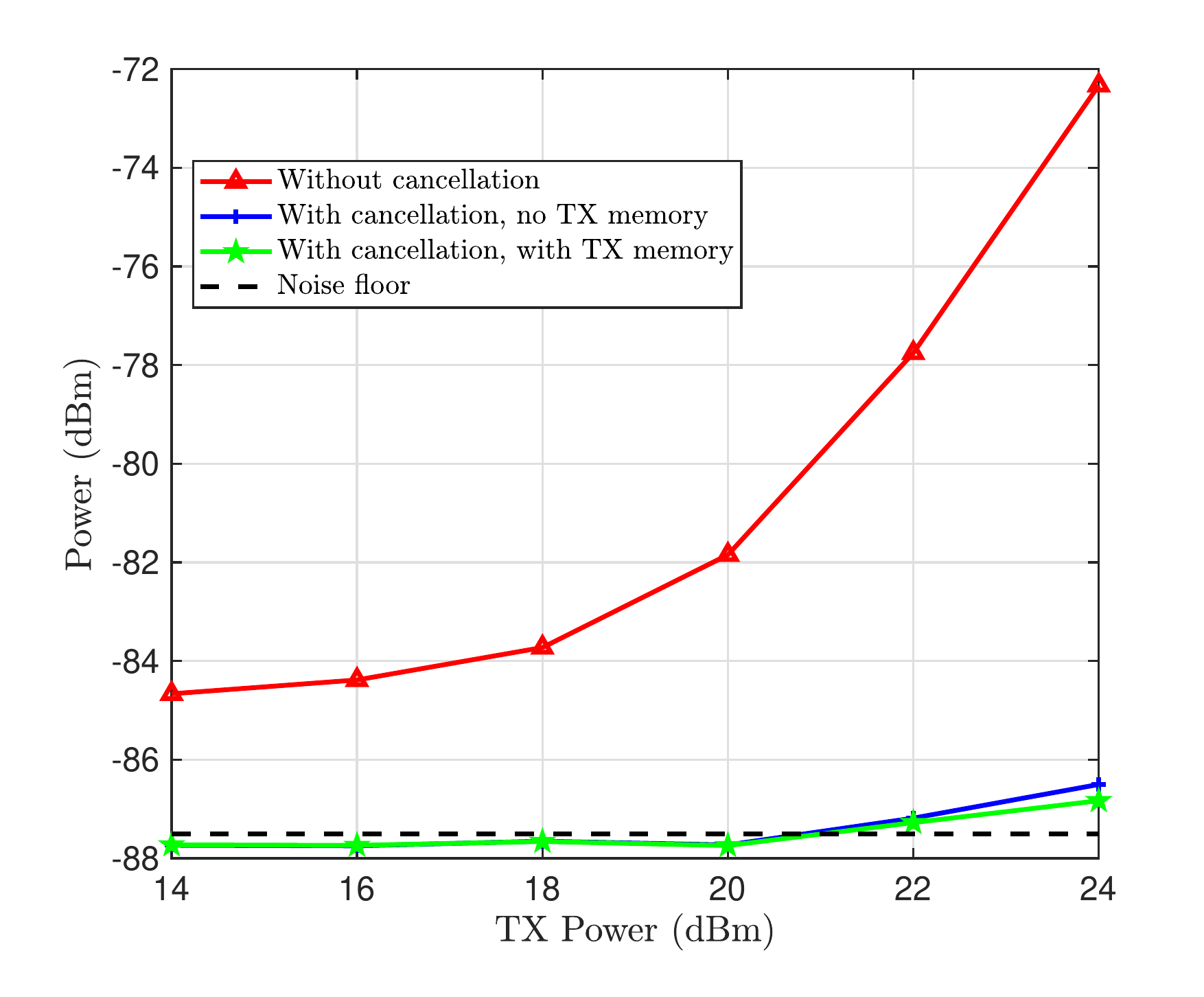}
\caption{{
The measured powers of the PIM distortion at the diversity RX branch of Band 1 without and with digital cancellation as functions of the aggregated transmit power. The TX CC bandwidths at the LTE Bands 1 and 3 are $5\;$MHz each.
}}
\label{CancCurvesdiversity_Rx}
\end{figure}

In this subsection, we continue the RF measurements and demonstrate that the PIM-induced distortion can also couple over-the-air and thus cause self-interference, e.g., to a diversity RX branch. To demonstrate this, a diversity RX antenna is placed at a distance of $5\;$cm from the main RX antenna, which provides an antenna isolation of around $10\;$dB. Fig.~\ref{diversity_Rx} shows then the observed PIM interference coupled over the air from the main transceiver into the diversity RX with the aggregated transmit power of $+24$ dBm. Again, the self-interference is substantial and clearly above the noise floor, and can thus degrade the receiver sensitivity. It can also be observed that the proposed digital cancellation approach is able to suppress the OTA-coupled self-interference efficiently very close to the noise floor. Finally, Fig.~\ref{CancCurvesdiversity_Rx} shows the PIM-induced distortion power at the diversity RX as a function of the aggregated TX power, evidencing that the proposed digital cancellation is able to efficiently suppress the self-interference close to the noise floor.

\subsection{Further Discussion}
While the presented results clearly demonstrate the modeling and cancellation capability of the proposed technique, it can be observed that there is still some residual interference present after the digital cancellation. This is particularly so when larger transmit powers are adopted. The main reason for this is that the developed cancellers consider only third-order PIM while neglect the higher-order distortion terms. The role of such higher-order components is more and more essential when the transmit power increases. Another reason is, despite elementary PA linearization, that also PA induced nonlinear distortion has an impact on the exact PIM waveform. 
Therefore, developing advanced techniques for joint compensation of PA-induced and PIM nonlinearities, with memory effects, and such that also higher-order nonlinear terms are taken into account \cite{Zeeshan_MTT_submit} forms an important research topic for our future work.

\section{Conclusion}\label{sec:conc}
In this paper, a novel digital cancellation solution was presented to estimate and cancel the self-interference resulting from the nonlinear passive components in CA-based radio transceivers. Along the self-interference modeling and the corresponding cancellation processing, the memory effects of the TX chains and PAs were also taken into account. The performance of the proposed digital cancellers was tested and analyzed with  actual RF measurements using real-life transceivers and RF components for UE devices, demonstrating excellent self-interference suppression. The presented measurement results also show that PIM can couple over the air into the diversity RX branch and thus cause significant interference in the diversity RX. Based on the obtained results, the proposed digital cancellation approach is an effective solution to suppress the PIM, both in the main RX as well as in the diversity RX chains. Such novel digital cancellation solutions can relax the linearity requirements of the passive RF components, while also enabling efficient utilization of the RF spectrum. Extending the developed cancellation solutions to accommodate higher-order nonlinear products as well as co-existing cascaded nonlinearities were identified as important future research topics.
\section*{Acknowledgment}
This work was supported by the Academy of Finland (under the grants 304147 and 301820), the Finnish Funding Agency for Innovation (Tekes, under the project ``5G Transceivers for Base Stations and Mobile Devices, 5G TRX''), Nokia Bell Labs, RF360, Pulse Finland, Sasken Finland, and Huawei Technologies Finland. The work was also supported by the Tampere University of Technology Graduate School.

\IEEEtriggeratref{5}

\end{document}